\documentclass[iop]{emulateapj}
 \usepackage{natbib}
\def\apjl{ApJ}
\def\apj{Astrophys. J.}

\def\mnras{MNRAS}

\def\aap{Astron. and Astrophys.}

\def\nat{Nature}

\def\pasp{PASP}
\def\aj{AJ}

\def\xmm{{\it XMM-Newton}}

\def\1023{PSR J1023+0038}
\def\ltsima{$\; \buildrel < \over \sim \;$}
\def\simlt{\lower.5ex\hbox{\ltsima}}
\def\gtsima{$\; \buildrel > \over \sim \;$}
\def\simgt{\lower.5ex\hbox{\gtsima}}
\def\es{$e^-/s\,$}

\received{}
\revised{\today}

\shorttitle{Kepler observations of \1023}
\shortauthors{Papitto et al.}

\begin{document}

\title{The first continuous optical monitoring of the transitional millisecond pulsar {\1023} with {\it Kepler}}

\author{A.~Papitto\altaffilmark{1}, N.~Rea\altaffilmark{2,3,4}, F.~Coti Zelati\altaffilmark{2,3},
  D.~de~Martino\altaffilmark{5}, S.~Scaringi\altaffilmark{6}, S.~Campana\altaffilmark{7}, E.~de O\'na Wilhelmi\altaffilmark{2,3},
  C.~Knigge\altaffilmark{8}, A.~Serenelli\altaffilmark{2,3}, L.~Stella\altaffilmark{1}, D.~F.~Torres\altaffilmark{2,3,9}, P.~D'Avanzo\altaffilmark{7},
  G.~L.~Israel\altaffilmark{1}}
\altaffiltext{1}{INAF--Osservatorio Astronomico di Roma, via Frascati
  33, I-00040, Monteporzio Catone (RM), Italy}
\altaffiltext{2}{Institute of Space Sciences (ICE, CSIC​)​, Campus UAB, Carrer de Can Magrans, 08193, Barcelona, Spain}
\altaffiltext{3}{Institut d'Estudis Espacials de Catalunya (IEEC), 08034 Barcelona, Spain}
\altaffiltext{4}{Anton Pannekoek Institute for Astronomy, University of Amsterdam, Postbus 94249, NL-1090-GE Amsterdam, The Netherlands}
\altaffiltext{5}{INAF--Osservatorio Astronomico di Capodimonte, Salita Moiariello 16, I-80131 Napoli, Italy}
\altaffiltext{6}{School of Physical and Chemical Sciences, University of Canterbury, Christchurch 8041, New Zealand}
\altaffiltext{7}{INAF--Osservatorio Astronomico di Brera, via Bianchi 46, I-23807 Merate (LC), Italy}
\altaffiltext{8}{School of Physics and Astronomy, University of Southampton, Highfield, Southampton SO17 1BJ, UK}
\altaffiltext{9}{Instituci\'o Catalana de Recerca i Estudis Avan\c{c}ats (ICREA), E-08010 Barcelona, Spain}

\begin{abstract}

  We report on the first continuous, 80 day optical monitoring of the
  transitional millisecond pulsar PSR J1023+0038 carried out in mid
  2017 with {\it Kepler} in the {\it K2} configuration, when an X-ray
  subluminous accretion disk was present in the binary. Flares lasting
  from minutes to 14 hr were observed for 15.6\% of the time, which is
  a larger fraction than previously reported on the basis of X-ray and
  past optical observations, more frequently when the companion was at
  superior conjunction of the orbit.  A sinusoidal modulation at the
  binary orbital period was also present with an amplitude of
  $\simeq$16$\%$, which varied by a few percent over timescales of
  days, and with a maximum that took place $890\pm85$~s earlier than
  the superior conjunction of the donor. We interpret this phenomena
  in terms of reprocessing of the X-ray emission by an asymmetrically
  heated companion star surface and/or a non-axisymmetric outflow
  possibly launched close to the inner Lagrangian point. Furthermore,
  the non-flaring average emission varied by up to $\approx40\%$ over
  a time scale of days in the absence of correspondingly large
  variations of the irradiating X-ray flux. The latter suggests that
  the observed changes in the average optical luminosity might be due
  to variations of the geometry, size, and/or mass accretion rate in
  the outer regions of the accretion disk.

\end{abstract}

\keywords{pulsars: individual (PSR J1023+0038) – stars: neutron – X-rays: binaries}

\section{Introduction} \label{sec:intro}

Binary systems hosting a neutron star (NS hereafter) that accretes
matter from a low mass donor (NS-LMXB) are the progenitors of
millisecond radio pulsars (MSP) powered by the rotation of the 
  pulsar magnetic field. The link between MSPs and NS-LMXBs has been
demonstrated by three transitional MSPs which switch between accretion
and rotation-powered states on time scales of days/months due to
variations of the mass inflow rate (PSR J1023+0038,
\citealt{archibald2009}; IGR J18245-2452, \citealt{papitto2013}; XSS
J12270-4859, \citealt{demartino2010,bassa2014}). In addition to
accretion outbursts ($L_X = 10^{36-37}$ erg s$^{-1}$) and
rotation-powered radio pulsar states ($L_X \approx 10^{32}$ erg
s$^{-1}$), all three known transitional MSPs have been observed also
into a peculiar X-ray sub-luminous disk state ($L_X = 10^{33-34}$ erg
s$^{-1}$).

\begin{figure*}[t!]
\includegraphics{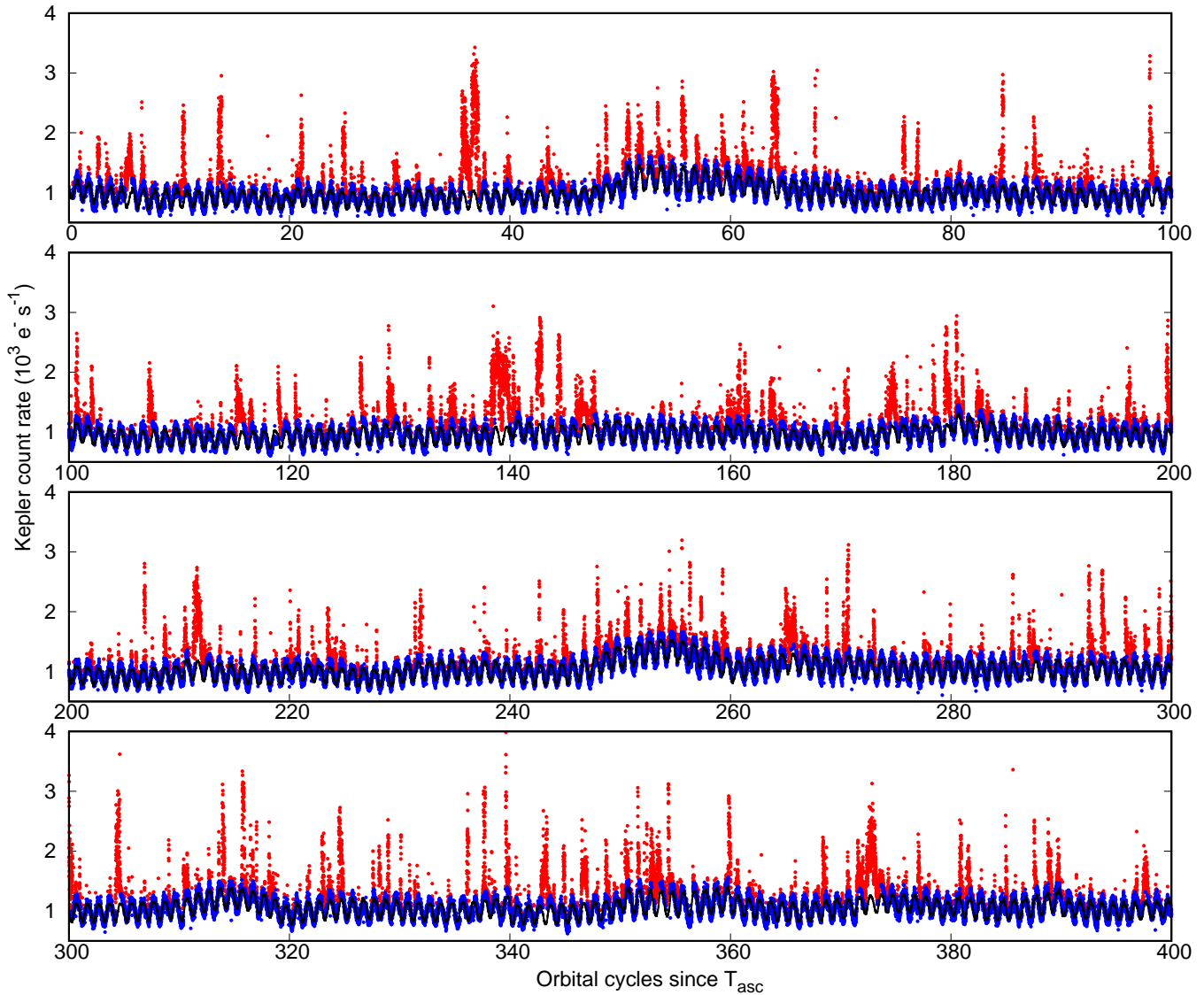}
\caption{ {\1023} optical light curve observed by {\it
    K2}. Time is expressed in orbital cycles since
  $T_{asc}=57905.1493086$~MJD, assuming $P_{orb}=17115.5216592$~s
  (\citealt{jaodand2016}, Papitto et al. in prep.). Flares are plotted
  with red dots, while the non-flaring emission is plotted using blue
  dots. The black solid line is the best-fitting model of the
  non-flaring emission (see text for details).}
\label{fig:lc}
\end{figure*}

Discovered as an eclipsing 1.69-ms radio pulsar in a 4.75-h binary
that previously had an accretion disk \citep{archibald2009}, \1023
entered in a sub-luminous disk state in 2013 June. The transition
featured the appearance of a double-peaked H$\alpha$ emission line in
the optical spectrum \citep{halpern2013}, the disappearance of radio
pulsations, and the increase of the X-ray and gamma-ray by a factor of
about ten and five, respectively \citep{stappers2014, patruno2014,
  tendulkar2014, torres2017}.  The X-ray emission switches from a {\it
  high} mode ($L_X\sim 7 \times 10^{33}$~erg s$^{-1}$; 0.3--79~keV,
70--80\% of the time), to a {\it low} mode ($L_X\sim 10^{33}$ erg
s$^{-1}$; 20\% of the time), and sometimes flares ($L_X\sim
5\times10^{34}$ erg s$^{-1}$, 2\% of the time; \citealt{bogdanov2015},
B15 hereafter, \citealt{jaodand2016}, J16 henceforth) on a timescale
of tens of seconds.  Coherent X-ray pulsations assumed to be
accretion-powered were detected during the {\it high} mode (but not in
the {\it low} mode; \citealt{archibald2015}). The optical emission of
{\1023} in the disk state ($g\simeq16.5$~mag) became one magnitude
brighter than in the radio pulsar state ($g\simeq17.5$), and was
dominated by the reprocessing of the X-ray emission from the disk and
the illuminated face of the donor \citep[][CZ14
  hereafter]{cotizelati2014}. Flares were also observed in the optical
band (\citealt{bond2002}, B15) and sometimes also a mode switching
similar to that observed in X-rays \citep[][S15
  henceforth]{shahbaz2015}.  Optical pulsations with an amplitude of
less than 1\% were also detected; even if observed when the source had
a disk, they could be hardly reconciled with accretion and suggested
that a rotation-powered pulsar might be operating
\citep{ambrosino2017}.  In this paper we present an extensive optical
monitoring performed with the NASA {\it Kepler} Space Telescope when
the source was in its current sub-luminous accretion disk state.

\section{{\it K2} observations, analysis and results}
\label{datakepler}

\begin{figure}[t!]
\includegraphics{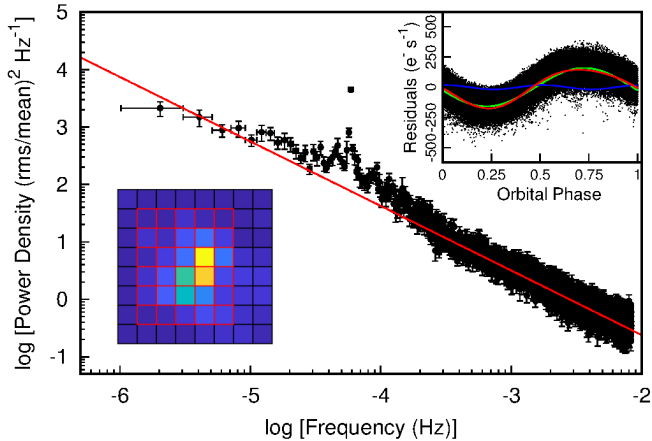}
\caption{Power density spectrum obtained averaging 15 intervals, each
  3 days-long, and normalized to give the squared rms fractional
  variability. The red-line shows a power law function
  $P(\nu)\propto\nu^{\beta}$, with $\beta=-1.12\pm0.01$. The inset in
  the lower left corner shows the {\it K2} average target pixel image
  for {\1023}. The target mask pixels are marked in red, whilst
  background mask pixels are marked in black.  The inset in the
    upper right corner shows the residual light curve folded at the
    4.75h orbital period, after removing the flares and the long-term
    trend evaluated with a quadratic function over 2 d-long
    intervals. The red solid line shows the best-fitting two-harmonics
    decomposition, the green and the blue solid lines show the first
    and the second harmonic, respectively. }
\label{fig:pds}
\end{figure}

{\1023} was observed by the {\it K2} mission during the campaign 14
between 2017-06-01 05:21:11 UT and 2017-08-19 21:56:19 UT (see
Fig.~\ref{fig:lc}). Here we analyse short cadence (SC) data (cadence
of 58.8~s) obtained from the Mikulski Archive for Space Telescope
(MAST) archive.  The data is provided in raw format, consisting
  of target pixel data. For each 58.8~s exposure we thus have a 8x8
  pixel image centered on the target.  {\1023} was near the edge of
  module 16.4. Although this module is not known to be affected by
  Moire Pattern Drift (MPD) noise, the target point spread function
  (PSF) is asymmetric as it lies at the edge of one of the outermost
  modules. As no other sources are known to exist within the 8x8 pixel
  images, we created the light curve by manually defining a large
  target mask as well as a background mask. A large target mask is
  required due to occasional small scale jittering of the spacecraft,
  resulting in the target moving slightly from its nominal position,
  as well as including the elongated PSF of {\1023}.  The inset in
the lower left corner of Fig.~\ref{fig:pds} shows the average masks
obtained from 117,030 individual target images, as well as the
  background pixels used.  We removed 1,691 observations because of
bad quality due to occasional spacecraft rolls or due to cosmic rays.
 Fig.~\ref{fig:pds} also shows the target and background masks in
  red and black respectively.  We produced the lightcurve by summing
together all target pixels for each exposure, and subtracted the
average background. The photometric time series was then 
  converted to the Solar System barycenter.

To convert {\it Kepler} flux to magnitudes, we considered two {\it
  Hubble Space Telescope} STIS/CCD 1 s-long unfiltered acquisition
(ACQ) images taken on 2017-06-13 and available at MAST
archive. Assuming the same spectral energy distribution observed in
2014 \citep[see Fig.~4.3 of][]{hern2016}, the {\it HST} count rate
translates into magnitudes of $g=16.47(4)$ and $r=16.48(4)$~mag. The
average flux observed by {\it Kepler} during the 1 minute-long
intervals overlapping with {\it HST} observations was $S_0=1323(5)$
\es; using the relation given by \citet{brown2011} to estimate the
Kepler magnitude as $K_p=0.1g+0.9r$ for $(g-r)\leq0.8$, we obtained
the conversion $K_p=16.48-2.5\log{(S/S_0)}$, where $S$ is the {\it
  Kepler} flux in \es. Variability on time scales shorter than the
Kepler time resolution and flux-dependent color changes both introduce
an uncertainty by $\sim0.1$~mag.

The {\it Kepler} light curve is highly variable ($<S>=1083$ \es,
$S_{rms}=286$ \es; Fig.~\ref{fig:lc}), reaching up to 3950 \es (i.e.,
$K_p\simeq$15.3~mag).  We identified: (i) flares lasting from minutes
up to $\sim14$ hours, (ii) a periodic modulation at the 4.75~hr binary
orbital period, and (iii) a variation of the average optical
luminosity over a time scale of days.  Figure~\ref{fig:pds} shows the
power density spectrum calculated averaging 3 days-long intervals, and
normalized to the fractional rms amplitude per unit frequency. The
spectrum is approximately described by a power law $P(\nu)\propto
\nu^\beta$ with $\beta=-1.12\pm0.01$ (errors are given at 1-$\sigma$
confidence level throughout the paper), similar to the value found by
S15, above which an excess peaking at the orbital frequency is evident
(see Fig.~\ref{fig:pds}).

\subsection{Flaring orbital variability}

\begin{figure}[t!]
\includegraphics{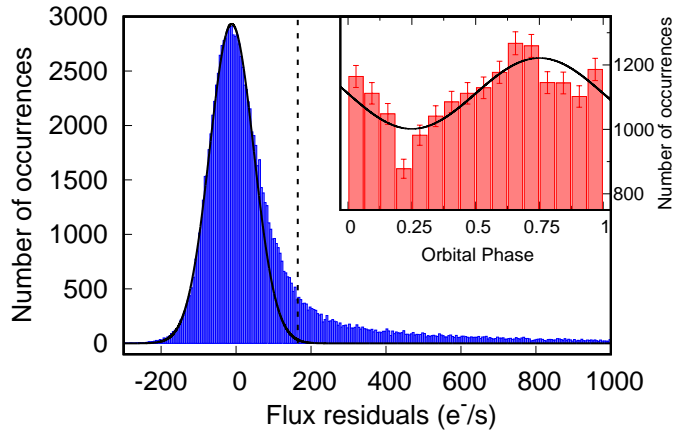}
\caption{Distribution of the flux residuals in $e^{-}/s^{-1}$ with
  respect to the best fitting function (see text for details),
  calculated over $\Delta t =58.8$~s-long time bins. The solid line
  represents the best-fitting Gaussian of the negative portion of the
  distribution. The dashed vertical line marks the threshold above
  which the source was identified as flaring. The inset shows the
  orbital phase distribution of the flares interval. Zero orbital
  phase corresponds to the passage of the pulsar at the ascending node
  of the orbit.  The best-fitting sinusoid with orbital period and
  phase fixed to the values derived from X-ray pulsar timing analysis
  is plotted as a black solid line. Maximum occurs at orbital phase
  0.75, i.e. when the donor star is at superior conjunction.}
\label{fig:basefluxdistro}
\end{figure}

In order to study the flare characteristics, we first isolated them
from the lower amplitude variations. To this aim we divided the {\it
  K2} light curve into $41$ intervals each spanning 10 orbital cycles
(i.e. $\simeq$2~days), and fitted the count rate with a function
consisting of the sum of a quadratic polynomial function and a Fourier
decomposition with period $P_{orb}=17115.5216592~s$ (J16). In some of
the intervals up to seven harmonic components were detected, while two
were enough in most of the cases.  We modeled with a Gaussian function
the negative portion of the distribution of light curve residuals with
respect to the best fitting function, and identified as flares the
$58.8$~s-long bins found above 3$\sigma$ the Gaussian median
value. Only the negative portion of the histogram was considered to
model the non-flaring flux distribution, as the positive end is
clearly contaminated by flares. Flares were then removed from the
light curve and the fitting procedure iterated until no further flares
were identified from the residuals. Figure~\ref{fig:basefluxdistro}
shows the final residuals histogram; time intervals with a residuals
larger than $\simeq165$~\es were deemed as flares. On average the
source flared for $15.6\%$ of the time covered by our {\it K2}
observations, a fraction variable between 6 and 30\% over $\simeq
2$~d-long intervals.  We fitted the 149 brightest flares with a
Gaussian profile determining the FWHM, time of maximum and
intensity. Flares appear of impulsive nature. They typically lasted
less than one hour with a distribution peaking around 45~minutes with
amplitudes spanning from 0.2 to 1~mag and no clear correlation between
duration and amplitude. When short they have a quasi-Gaussian shape
whilst the longer ones appear to be  made up by trains of short
impulsive flares.

To study the orbital phase dependence of the flares, we evaluated the
orbital phase of each flaring time bin using the epoch of passage of
the pulsar at the ascending node determined from the X-ray pulsations
detected in the {\xmm} observations performed on 2017, May 23 and 24,
$T_{asc}=57896.8292633(2)$~MJD (Papitto et al., in preparation) as the
zero phase reference. Note that in these units the inferior
conjunction of the donor star takes place at phase 0.25.  The flare
orbital phase histogram has a maximum around phase $0.75$ (see the
inset of Fig.~\ref{fig:basefluxdistro}), i.e. when the donor star is
at superior conjunction.  The fractional amplitude evaluated from a
sinusoid with fixed period and phase is $A=(9.9\pm2.0)$\%.  
We note that the bin time adopted to study the orbital dependence of
flare occurrence is such that the longest flares weighted more in the
overall phase distribution.

\subsection{Non-flaring orbital variability}

\begin{figure}[t!]
\includegraphics[width=8.6cm]{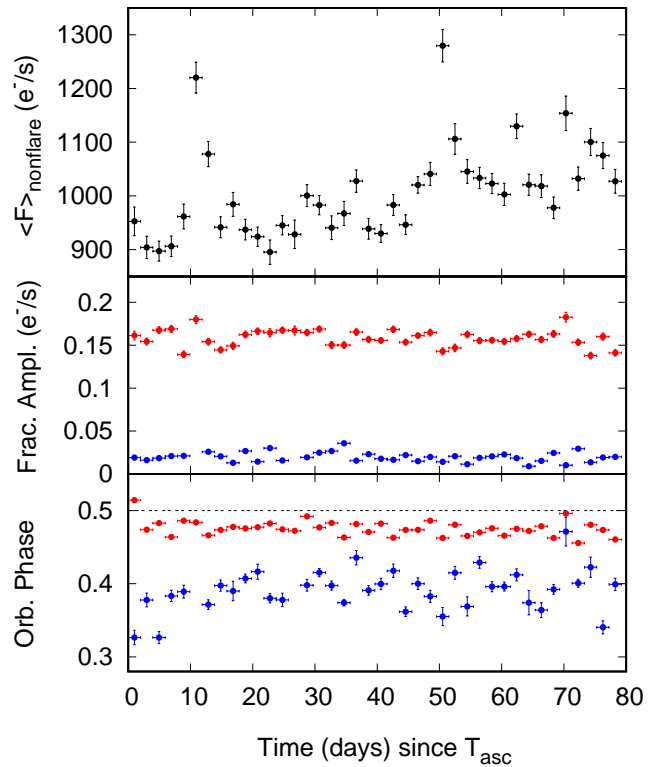}
\caption{Non-flaring average flux evaluated as the mean of the
  quadratic function used in fitting time intervals of the {\it
    Kepler} light curve spanning 10 orbital cycles (i.e., 47.5~h,
  top panel). Fractional amplitude of the first (red dots) and second
  (blue dots) harmonic of the modulation at the orbital period of the
  system (middle panel). Phase of the first (red dots) and second
  (blue dots) harmonic with respect to the epoch of passage of the
  pulsar at the ascending node according to the X-ray pulsar timing
  solution (bottom panel).  }
\label{fig:trend}
\end{figure}

A bimodality of the non-flaring emission between a {\it high} and a
{\it low} mode each lasting for tens of minutes, similar to that
observed in X-rays (B15) and reported in the optical band by S15
(compare their Fig.~4), did not appear neither in the histogram of
residuals calculated over the whole {\it K2} exposure
(Fig.~\ref{fig:basefluxdistro}), nor those ranging over
2~days-intervals.  Restricting the analysis to 2 orbital cycles,
  hints of a bimodality appeared in a dozen out of the 200 intervals
  considered, even if sharp-edged rectangular dips could be hardly
  detected in the residual light curve.  In those intervals, the {\it
    low} mode occurred more frequently than the {\it high} one,
  similar to what was found by S15 and opposite with respect to the
  X-ray behaviour.  Fig.\ref{fig:trend} shows the trend of the
non-flaring flux evaluated as the average value of the quadratic
function used in fitting time intervals spanning $\sim$2~days, the
fractional amplitude and the phase of the first and second harmonic
employed to fit the orbital modulation. The average non-flaring flux
was clearly variable. Four episodes of brightening can be identified
with count rate increasing by up to $40$\% on a time-scale of
$\simeq4$~days.  There was no correlation between the average
non-flaring flux and the total fluence of flares. The two peaks at
$\approx 10$ and $\approx 50$ days since $T_{asc}$ might hint at a
super-orbital variability, but the interval covered by {\it K2} data
is too short to assess that.  The fractional amplitude of the first
and second harmonics used to model the orbital modulation are variable
by a few percent around an average value of $<A_1>=15.9\pm0.1$\% and
$<A_2>=1.8\pm0.1$\%, respectively.  The average phases of the first
and second harmonic are $<\phi_1>=0.476(2)$ and $<\phi_2>=0.393(4)$,
respectively; as a consequence the maximum of the sinusoidal
modulation takes place on average at phase $<\phi_{max}>=0.698$ (with
a standard deviation of 0.031), $890\pm85$~s earlier than the donor
superior conjunction.   The average orbital modulation is plotted
  in the inset in the upper right corner of Fig.~\ref{fig:pds}, as the
  light curve residuals folded at the 4.75h orbital period, after
  removing the flares and the long-term trend evaluated with a
  quadratic function over 2 d-long intervals.

\section{Discussion}
\label{discussion}

The multiwavelength phenomenology of the transitional millisecond
pulsar {\1023} in the sub-luminous accretion disk state is exteremely
complex. Variable emission on time-scales ranging from millisecond
to years was observed from the radio up to the gamma-ray energy
bands. The analysis of the uninterrupted {\it K2} light curve
presented here gives an unprecedented view of its the optical
variability from minutes to 80 days.

More than $170$ optical flares were observed (i.e.,
$\simeq2$~day$^{-1}$), both short ($\sim$ minutes) and long (up to
$\sim14$ hours), with no significant correlation between the duration
and the amplitude. The brightest flares attained $K_p=15.3$~mag,
i.e. $\simeq 1.25$~mag brighter than the non-flaring emission. While
small samples of optical flares with similar properties have been
already reported by \citet{bond2002}, B15, S16, and \citet[][HK18
  henceforth]{hakala2017}, the long uninterrupted {\it Kepler} coverage
allowed us to determine for the first time that on average the source
was flaring $15.6\%$ of the time, with a flare occurrence of 20\%
higher at orbital phases corresponding to the donor superior
conjunction.

Flares from {\1023} were observed simultaneously in the X-ray, UV,
optical and NIR bands, suggesting they are related to the same process
(B15, HK18), with most of the energy released at X-ray energies
($L_X\simeq6\times10^{34}$~erg~s$^{-1}$, $L_{X}/L_{opt}\simeq 6$). We
compare the fraction of the {\it K2} light curve characterized by
optical flares to that observed in the X-ray band, where {\1023} has
been intensively observed. The {\it Swift} satellite observed {\1023}
for 30 times simultaneously with the {\it Kepler} campaign, for a
total on-source time of $\sim$1.1~days, but the relatively low photon
statistics of the {\it Swift} X-ray Telescope only allowed us to
identify clearly one X-ray flare associated to an optical flare. On
the other hand, {\1023} was observed 10 times by {\xmm} for a total
exposure of 5.6~days between 2013 June and 2015 December \citep[][B15,
  J16]{archibald2015}. The source spent $\simeq2\%$ of the time
flaring in X-rays (J16), less than the flaring time fraction observed
by {\it K2} even taking into account its variability over 2~days-long
intervals (values as low as 6\% were observed). Only part of this
discrepancy can be ascribed to the lower count rate recorded by {\xmm}
with respect to {\it Kepler}. X-ray flares were in fact identified by
{\xmm} when the ratio between the 0.5--10~keV net flare count rate and
that observed in the {\it high} mode was larger than
$(S_{flare,min}/S_{high})_{Xray}\simeq 11/7-1\simeq 0.6$ (B15). For
the brightest flares, the ratio $\xi$ between the flare and
non-flaring flux observed in the optical ($(S_{flare}/<S>)_{opt}\simeq
4000/1000 - 1\simeq3 $) and in the X-ray band
($(S_{flare,max}/S_{high})_{Xray}\simeq 60/7-1\simeq 7.5$, see
e.g. Fig.~7 in J16) was $\xi\simeq3/7.5=0.4$. Assuming that this ratio
held for all flares - a conservative hypothesis as not all the X-ray
flares have an optical counterpart - we conclude that the fainter
flares detectable by {\xmm} would have a residual {\it Kepler} flux of
$\xi\times (S_{flare,min}/S_{high})_{Xray} \times <S>_{Kep}\simeq
260$~\es. The fraction of time spent by {\1023} during {\it K2}
observations above such a threshold is $10.3\%$. We conclude that even
taking into account variability on intervals of a few days and
difference in the sensitivity of {\it Kepler} and {\xmm}, the observed
flaring time fraction observed by {\it Kepler} can be taken as an
indication that the source spent flaring a larger portion of the time
in June/July 2017 than in 2013-2015. It is not clear what may cause
such an increase, as the X-ray properties observed by {\it Swift}
simultaneously to the {\it K2} campaign 14, and during {\xmm}
observations performed on 23 and 24 May, 2017, i.e. just before it,
were consistent with those determined during previous years.

Our analysis showed that optical flares were seen more often close to
the donor superior conjunction, whereas no dependence of the X-ray
flare detection on the orbital phase has yet been reported, possibly
because only a handful of X-ray flares have been actually observed so
far.   Note that, however, after this paper was submitted we
  became aware of an analysis of the same {\it K2} data set by
  \citet{kennedy2018}, who used a slightly different algorithm to
  define flares. According to their definition flares occurred more
  often than from our analysis (22\% of the time) and the flare peaks
  did not show a significant dependence on the orbital phase. We then
  conclude that the orbital dependence of flares is uncertain and
  heavily depends on their definition. If confirmed, a dependence of
the optical flares observation rate on the orbital phase would
indicate that at least part of the flares originated from reprocessing
of the X-ray emission off the surface of the companion star.  However,
the spectral colors observed by S15 from one bright optical flare
indicated that it originated in an optically thin medium such an
accretion disk corona and/or hot fireball ejecta. These could be
launched by a propellering magnetosphere
\citep{papitto2014,papitto2015}, or by the wind of a rotation-powered
pulsar assumed to be active in spite of the presence of an accretion
disk (\citealt{takata2014}, CZ14, \citealt{ambrosino2017}). In the
latter case, flares could be produced by the interaction of a
relativistic pulsar wind with clumps of matter in the region beyond
the light cylinder \citep[see, e.g.][]{zdziarksi2010}. A change of the
accretion flow during flares possibly related to absorption by ejecta
of matter is also supported by the lack of H$_{\alpha}$ emission from
the portion of the disk that is closest to the observer in between
orbital phases 0.25--0.5, and by the appearance of an additional
polarised emission (HK18; see also \citealt{baglio2016}). Assuming
that flares originated from the ejecta, an orbital dependence could
indicate that matter is preferentially launched from a spot in the
disk along the line that joins the pulsar and the donor star, possibly
close to the inner Lagrangian point. A similar scenario was put
forward by \citet{demartino2014} to explain the disappearance of
emission lines orginated in the outer disk regions of the transitional
MSP XSS J12270--4859, when the donor was at its superior conjunction.

The first and second harmonics of the Fourier decomposition used to
model the orbital modulation had an average fractional amplitude of
$\simeq 15.9$ and $\simeq 1.8$\%, respectively. Generally, the first
harmonic is ascribed to irradiation of the companion star by the X-ray
source with an expected maximum close to the donor superior
conjunction, while the second is assumed to trace the ellipsoidal
deformation of the companion star and has maxima when the donor is at
quadrature.  On average, the observed orbital modulation had a maximum
that  preceded the donor superior conjunction by
$890\pm85$~s. Assuming that most of the orbital modulation arises from
the vicinity of the inner Lagrangian point, and taking
$M_1=1.4$~M$_{\odot}$ and $M_2=0.2$~M$_{\odot}$ for the mass of the
primary and the donor, respectively, this lag translates into an
arc-length of $\simeq2.5\times10^9$~cm, which is smaller than the
radius of the donor star ($R_2\simeq 3\times10^{10}$~cm). A number of
effects can produce the observed lags between the phase of the first
harmonic of the optical orbital modulation and the donor superior
conjunction; namely (i) asymmetric heating of the companion due to
screening of irradiating X-rays by intervening material such as matter
ejected from the disk, (ii) a contribution from matter streaming past
the inner Lagrangian point, (iii) absorption of the companion star
emission by the ejecta launched from the vicinity of the inner
Lagrangian point.

The average non-flaring optical emission varied by up to ~40\% over
timescales of a few days during the {\it K2} campaign, ranging from
$K_p=16.9$ to $16.5$.  In comparison the average flux reported by CZ14
corresponds to a {\it K2} magnitude of $K_p=16.36(2)$. Similar large
variations of the X-ray flux were not observed in past observations,
nor in the {\it Swift} monitoring during the {\it Kepler}
observations. B15 reported that the X-ray flux observed in the {\it
  high} mode (in which the source spends $\sim70$\% of the time)
decreased by just a few per cent between 2013 November and 2014
June. In addition, the X-ray flux observed in the {\it high} and {\it
  low} mode during {\xmm} observations performed one week before the
start of the {\it K2} campaign was still consistent with that
determined by B15 about three years earlier.  We did not found a
correlation between the optical flare fluence (possibly related to the
fluence of the X-ray flares) and the average non-flaring optical
emission. The observed variations of the non-flaring optical emission
could be then due to variability of the outer disk intrinsic emission
caused by variations in the mass in-flow rate, that would not be
reflected by a concurrent increase in the X-ray emission possible
because of mass ejection before the mass gets close enough to the NS
to emit high energy radiation. Alternatively, changes in the angle
subtended by the medium that reprocesses high energy radiation into
visible emission could have occurred, suggesting a very complex and
variable geometry of the accretion flow between the inner Lagrangian
point and the outer disk regions, and/or azimuthal extent changes of
the latter.

Finally we note that hints for a bimodal flux distribution like those
reported at some epochs by S15  appeared only in a dozen 2 orbital
  cycles-long intervals (i.e., $\sim 8\%$ of the total) of the {\it
  K2} data we analysed.  Optical dips were  not detected at other
epochs (J16, HK18), indicating that these features are not stable as
in the X-ray band. When observed (S15), dips had ingress/egress times
of 12-35~s, duration in the range 80--1300~s, and a separation
200--1900~s. These properties are similar to those of the {\it low}
mode dips observed in the X-ray band for $\sim20\%$ of the time (B15,
J16), and were claimed to be their equivalent, even if simultaneous
observations were lacking.  Note that when a bimodal distribution
  of the optical flux is observed (S15, or in a few intervals of the
  {\it K2} data analysed here, the low flux mode is predominant,
  whereas the opposite happens in the X-rays. The time resolution
($\sim$60s) of {\it K2} data could well cause smearing of the sharp
ingress/egress shape of the dips but should not have prevented the
detection of the long (20~min) dips interleaved by less than an
hour.  Even when a bimodality of the flux was suggested by the
  residual flux distribution, the light curve did not show such
  prolonged dips.
 Future simultaneous optical and X-ray monitoring will
possibly ascertain whether the lack of optical dips marked a change in
the properties of the source.

\acknowledgements

A.P. acknowledges funding from the EU's Horizon 2020 Framework
Programme for Research and Innovation under the Marie
Sk{\l}odowska-Curie Individual Fellowship grant agreement
660657-TMSP-H2020-MSCA-IF-2014, and from the agreements ASI-INAF
I/037/12/0 and ASI-INAF 2017-14-H.0. N.R., F.C.Z., and
D.F.T. acknowledge funding from grants AYA2015-71042-P and SGR
2017-1383.  A.S. acknowledges funding from grants ESP2017-82674-R
(MINECO) and 2017-SGR-1131.


\end{document}